# Choreographies for Automatic Recovery*

Claudio Antares Mezzina[1] and Emilio Tuosto[2]

[1] IMT School for Advanced Studies Lucca, Italy
claudio.mezzina@imtlucca.it
[2] Department of Informatics, University of Leicester, Leicester LE1 7RH, UK
emilio@le.ac.uk

**Abstract.** We propose a choreographic model of reversible computations based on a conservative extension of global graphs and communicating finite-state machines. The main advantage of our approach is that does not require to instrument models in order to control reversibility but for a minor decoration of branches. We show that our models are conservative extensions of existing ones and that the reversible semantics guarantees causal consistency.

## 1 Introduction

To make your app more effective you would like it to dynamically select an external service so that, when the current service in use becomes slow, a new service is tried. Of course such an adaptive behaviour could be explicitly programmed. Or should it? Another way to attain such behaviours could be to use *reversible computations*. Reversible computing is attracting interest as a suitable abstraction for a range of application domains (e.g., transactions and fault tolerant systems). The ability to potentially undo computations provides us with an ideal setting to study, revisit, or imagine alternatives to standard techniques for building dependable systems as well as to debug them. Current models either focused on studying the interplay of reversibility and concurrency in calculi [3, 13, 9] or have some limitations that we discuss in Section 7.

We propose a reversible choreographic model as a suitable way of smoothly encompassing reversible computation in the specification of communication-centric applications. The basic idea leverages on the so-called *top-down* approach of choreographies for distributed applications, whereby one specifies a global model, the so-called *global view*, from which automatically *projects* distributed components interacting via message-passing, the so-called *local view*. This is a main benefit of choreographic approaches which allow designers to abstract away from low level details of the underlying distribution and communication models.

We adopt *global graphs* [4, 10, 6] to formalise global views of choreographies and *communicating finite-state machines* (CFSMs) [2] to formalise local views of choreographies. Global graphs can be seen as a graphical model of distributed work-flows and communicating machines (a well established model for communication protocol design) fix the distribution and communication model. In passing

* Research partly supported by the EU COST Action IC1405.

we note that the model of CFSMs is very close to the actor model, which is recently gaining momentum in mainstream programming languages.

Following the principle of choreographies as lightweight abstractions, we study a linguistic modification of global graphs to support reversibility of distributed interactive components. More precisely, we propose a simple conservative extension of global graphs and CFSMs which allows designers to specify conditions to revert branches of distributed choices. The extension of global graphs we propose, despite its simplicity, has a deep impact on the operational semantics of the communicating machines. In fact, we have to instrument the standard semantics of communicating machines with new elements to keep track of the execution in order to establish when it can be reverted. A key advantage of our approach is that the coordination required when reversing computations can be automatically derived by projecting global graphs to CFSMs.

*Main contributions* Reversible global graphs (cf. Section 2) are defined by simply decorating branches of a choice with a *reversion guard* $\phi$, that is with a condition $\phi$ on the state of the system triggering a backward computation. The idea is that the designer specifies distributed choices so that, when a branch selected for execution, the computation proceeds forward until the guard of the branch becomes true. In this case, the run-time tries to revert the computation in order to find a "better" branch to execute, if such branch exists; otherwise the initially selected branch is completed.

This minor extension of global graphs requires a rather deep modification of the model of local views: for this we propose *reversible* CFSMs (cf. Section 3). The idea is to maintain the information about history of computation in the configurations of systems of CFSMs. In this way, at run-time we can establish what communications can be reversed and how to bring back the system to a previous state. Interestingly, we do not resort to any notion of reversible automaton [11], that is automata where transitions may have explicit reversed transactions; this makes our r-CFSMs just CFSM machines carrying extra information.

The forward semantics of reversible CFSMs (cf. Section 4) is defined according to the same principles of the original semantics [2]. Interestingly, our constructions yield a conservative extension of existing models. In fact, the semantics of reversible global graphs is basically borrowed from [6] and simply ignores reversion guards, while Theorem 1 and Theorem 2 show that the forward semantics of reversible CFSMs is just a decoration of the operation semantics of CFSMs.

The projections of reversible global graphs on reversible CFSMs (cf. Section 5) is also similar to the corresponding construction in [6]. The main difference is that we have iterative choreographies and their interplay with reversible computations. This requires to explicitly mark when loops start and finish.

Finally, the backward semantics (cf. Section 6) builds on the previous construction in order to identify the points in the configuration of a system where communications can be reversed. We note that this is done at run-time and does not require explicit primitives (i.e., reversibility is controlled by the semantics [8]). The adequacy of our backward semantics is shown by Theorem 3.

## 2 Reversible Global Graphs

Let $\mathcal{P}$ be a finite set of *participants* (ranged over by $\mathsf{A}$, $\mathsf{B}$, etc.), $\mathcal{M}$ a set of *messages* (ranged over by $\mathsf{m}$, $\mathsf{x}$, etc.), and $\mathbb{Z}_0$ the set of integers different from 0. We take $\mathcal{P}$, $\mathcal{M}$, and $\mathbb{Z}_0$ pairwise disjoint. The participants of a choreography use *channels* to exchange messages and coordinate with each other. Formally, a channel is an element of the set $\mathcal{C} \triangleq \{(\mathsf{A}, \mathsf{B}) \in \mathcal{P} \times \mathcal{P} \mid \mathsf{A} \neq \mathsf{B}\}$ and we let $\mathsf{A}{\cdot}\mathsf{B}$ to range over $\mathcal{C}$. The next definition introduces a variant of global graphs that extends the language in [6] with iteration and generalises the branching construct to account for reversibility.

**Definition 1 (Reversible choreography).** *The set $\mathcal{G}$ of* reversible choreographies *(r-choreographies for short) consists of the terms $\mathsf{G}$ derived by the grammar*

$$\begin{aligned}
\mathsf{G} \;::=\;& \mathsf{A} \xrightarrow{i} \mathsf{B} \colon \mathsf{m} && (1)\\
\mid\;& \mathsf{G};\mathsf{G}' && (2)\\
\mid\;& i.(\mathsf{G} \mid \mathsf{G}') && (3)\\
\mid\;& i.\big({*}\mathsf{G} @ \mathsf{A}\big) && (4)\\
\mid\;& i.\big(\mathsf{G}_1 \ \mathtt{unless}\ \phi_1 + \cdots + \mathsf{G}_h \ \mathtt{unless}\ \phi_h\big) && (5)
\end{aligned}$$

*that satisfy the following conditions:*

- $0 < i \in \mathbb{Z}_0$ *is an integer called* control point
- *any two control points occurring in different positions of a choreography are different (e.g.,* $1.(\mathsf{A} \xrightarrow{2} \mathsf{B} \colon \mathsf{m} \mid \mathsf{C} \xrightarrow{1} \mathsf{D} \colon \mathsf{y}) \notin \mathcal{G}$*)*
- *in* (1), $\mathsf{A}{\cdot}\mathsf{B} \in \mathcal{C}$
- *in* (4), $\mathsf{A}$ *is a participant of* $\mathsf{G}$ *(e.g.,* $i.\big({*}(\mathsf{A} \xrightarrow{j} \mathsf{B} \colon \mathsf{m}) @ \mathsf{C}\big) \notin \mathcal{G}$*).*
- *in* (5), $\phi_1, \ldots, \phi_h$ *are* reversion guards *(logical formulae defined below).*

*For* $\mathsf{G} \in \mathcal{G}$*, let* $\mathsf{cp}(\mathsf{G})$ *denote the set of control points in* $\mathsf{G}$.

An r-choreography can be a simple interaction (1), the sequential (2) or parallel (3) composition of r-choreographies, the iteration of a choreography (4), or the choice between several r-choreographies (5).

Control points $\mathsf{i}$ tag static positions of r-choreographies where a communication happens (cf. (1)) or the execution flow changes (due to parallelisation, or iteration, or non-deterministic choices). In the following, we may omit control points when immaterial, e.g., writing $\mathsf{G} \ \mathtt{unless}\ \phi + \mathsf{G}' \ \mathtt{unless}\ \phi'$ instead of $\mathsf{i}.\big(\mathsf{G} \ \mathtt{unless}\ \phi + \mathsf{G}' \ \mathtt{unless}\ \phi'\big)$. Also, the actual values of control points are irrelevant and therefore we consider equivalent r-choreographies that differ only on the values of control points.

Iteration (4) can intuitively be thought of as repeat-until loop where participant $\mathsf{A}$ decides whether to repeat the body $\mathsf{G}$ or exit the iteration. Note that the body $\mathsf{G}$ is executed at least once. The new branching construct (5) generalises the usual branching construct of choreographies so to control reversible computations with the following intended meaning: the choreography $\mathsf{G}_i$ of a non-deterministically chosen branch $1 \leq n \leq h$ is executed until the guard $\phi_n$ remains false; if instead $\phi_n$ becomes true during the execution of $\mathsf{G}_n$ **and** the part of $\mathsf{G}_n$ executed so far can be reversed, then another of the not yet tried branches is chosen, **if any**. When all branches have been tried once or all the guards are true, a branch is non-deterministically chosen and fully executed without further backtracking. Our framework is parametric wrt to the chosen language of guards. For the sake of this paper, we consider a simple propositional logical language $\Phi$ of guards; to fix the notation we will assume that $\Phi$ consists of the formulae generated by the following grammar:

$$\phi \ ::= \ \mathtt{p}(\mathsf{A}_1 \cdot \mathsf{B}_1, \cdots, \mathsf{A}_\mathsf{N} \cdot \mathsf{B}_\mathsf{N}) \ \big| \ \mathsf{m} \in \mathsf{A} \cdot \mathsf{B} \ \big| \ \neg \phi \ \big| \ \phi \vee \phi$$

where $\mathtt{p}$ is an $n$-ary predicate taken from a set of symbols that we leave unspecified. Basically, guards can be thought of as propositions predicating on channels and messages. In examples, we allow ourselves to use eg numerical constants in guards and we let $\mathtt{tt}$ and $\mathtt{ff}$ to be the usual truth values 'true' and 'false' respectively.[3]

Through the paper we will consider a running example where a travel agent T repeatedly queries a broker B for the price of a flight and of a car separately or altogether. Once the price(s) are returned, T updates its database D.

*Example 1.* The protocol of our running example can be expressed by the following choreography

$$\mathsf{G}_{(6)} = 1.\big(*\big(2.\big(\mathsf{G}_1 \ {}_{\text{unless}} \ \phi_{(6)} + \mathsf{G}_2 \ {}_{\text{unless}} \ \phi'_{(6)}\big);\mathsf{T} \xrightarrow{10} \mathsf{D}\colon \mathsf{upd}\big)@\mathsf{T}\big) \qquad (6)$$

where

$$\mathsf{G}_{1(6)} \triangleq 3.(\mathsf{T} \xrightarrow{4} \mathsf{B}\colon \mathsf{flight};\mathsf{B} \xrightarrow{5} \mathsf{T}\colon \mathsf{flightPrice} \,|\, \mathsf{T} \xrightarrow{6} \mathsf{B}\colon \mathsf{car};\mathsf{B} \xrightarrow{7} \mathsf{T}\colon \mathsf{carPrice})$$
$$\mathsf{G}_{2(6)} \triangleq \mathsf{T} \xrightarrow{8} \mathsf{B}\colon \mathsf{dest};\mathsf{B} \xrightarrow{9} \mathsf{T}\colon \mathsf{fullPrice}$$

The guards are defined[4] as $\phi'_{(6)} \triangleq \neg\phi_{(6)}$ and $\phi_{(6)} \triangleq |\{\mathsf{upd} \in \mathsf{T}\cdot\mathsf{D}\}| < 1$ which reads as "the number of upd messages in the channel T·D is below 1". □

The syntax in Definition 1 captures the structure of a visual language of directed graphs so that each r-choreographies G can be represented as a rooted graph with a single "enter" and a single "exit" control points called *source* and *sink* respectively. Fig. 1 illustrates our graphical notation. There, nodes ○ and ◎ respectively denote the source node the sink node; other nodes are drawn as • and a dotted edge from/to a •-control points singles out the source/sink control point the edge connects to. For instance, in the graph for the sequential composition, the top-most edge identifies G sink node and the other edge identifies the source node of G′; intuitively, • is the control point of the sequential composition of

[3] Truth values can be derived in the obvious way, eg $\mathtt{tt}$ is a macro for $\neg\phi \vee \phi$.
[4] Guards are application-dependent and could be completely unrelated in general. In our running example we take $\phi'_{(6)}$ to be the negation of $\phi_{(6)}$ just for simplicity.

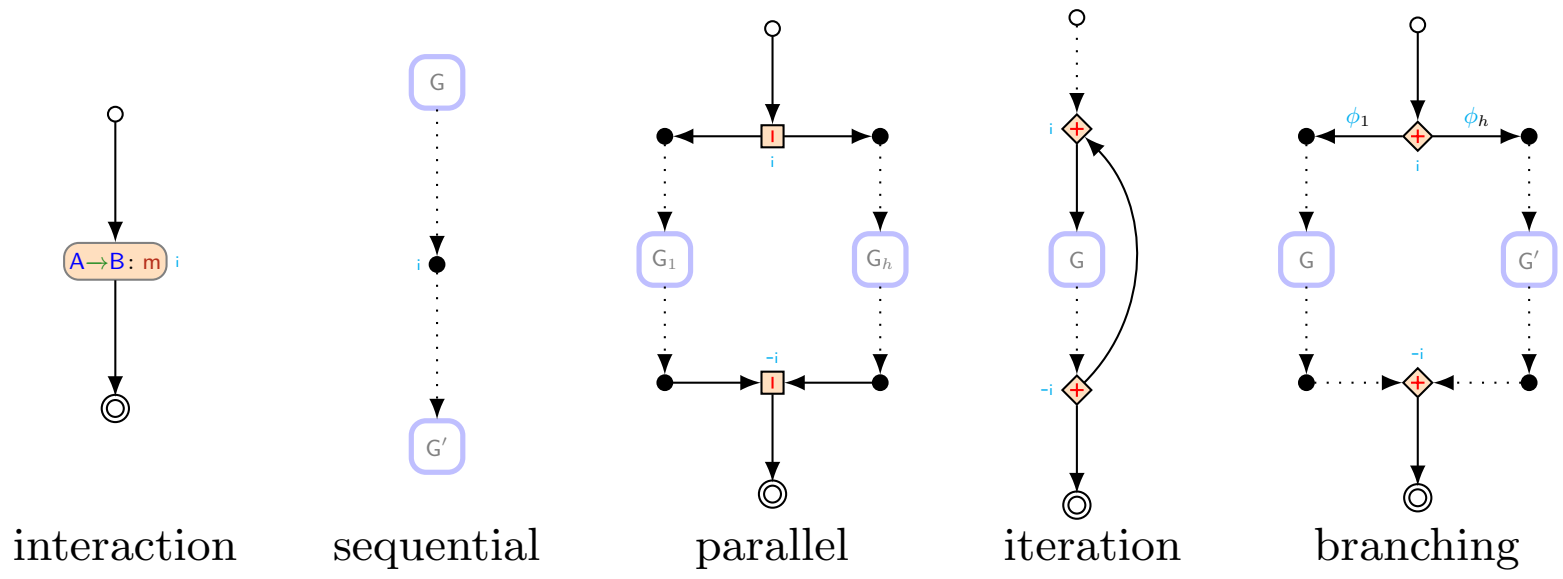

**Fig. 1.** A graphical notation for r-choreographies

G and G′ obtained by "coalescing" the sink control point of G with the source control point of G′. In a graph $\mathsf{G} \in \mathcal{G}$, each control point i marks either a branch or a fork gate, respectively graphically depicted as ◈ and ▣; to each branch/fork control point also corresponds a "closing" control point -i marking the merge/join point of the execution flow. Labels will not be depicted when immaterial.

*Example 2.* The graphical representation of the global graph of our running example is

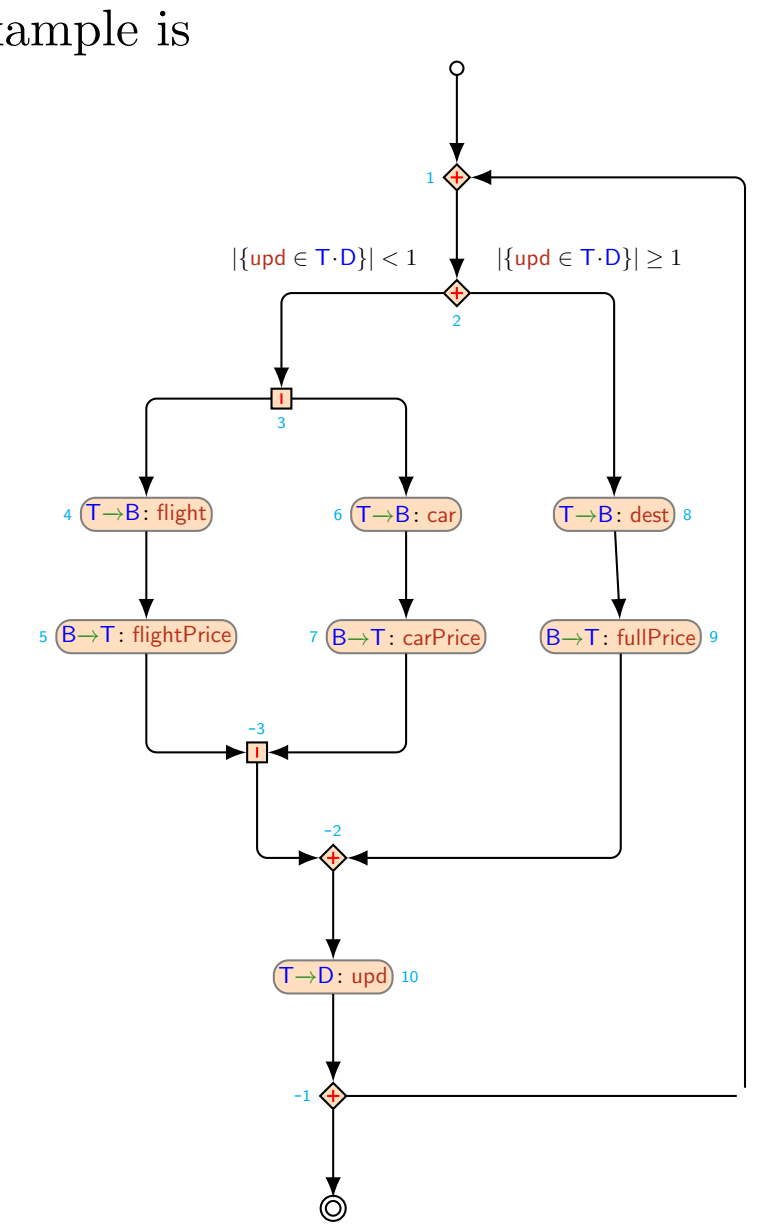

□

The semantics of r-choreographies is obtained from the one in [6] by simply ignoring guards.[5] For space limitations, we give only an intuitive idea of such semantics and refer the reader to [6] for the details. The semantics $[\![\mathsf{G}]\!]$ of a choreography $\mathsf{G} \in \mathcal{G}$, when defined, induces a partial-order on the communication

[5] Guards will be used in Section 3 to reverse computations of CFSMs.

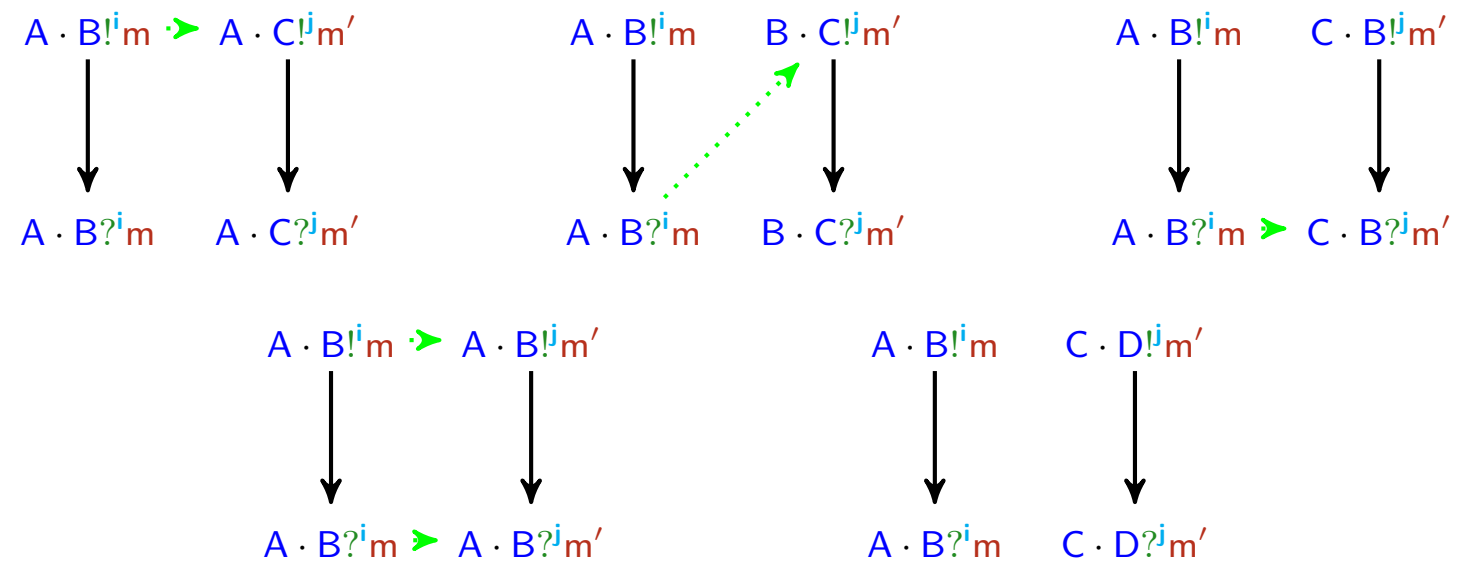

**Fig. 2.** Examples of sequential composition

*events* of $\mathsf{G}$, denoted by $\leq_{\mathsf{G}}$. Such events are formalised as elements of the set $\mathcal{E}$ of *events* ranged over by $e$, $e'$, ... and defined by

$$\mathcal{E} \triangleq \mathcal{E}^! \cup \mathcal{E}^? \qquad \text{where} \qquad \mathcal{E}^! = \mathcal{C} \times \{!\} \times \mathbb{Z}_0 \times \mathcal{M} \quad \text{and} \quad \mathcal{E}^? = \mathcal{C} \times \{?\} \times \mathbb{Z}_0 \times \mathcal{M}$$

Events in $\mathcal{E}^!$ and $\mathcal{E}^?$ respectively represent *sending* and *receiving* events; $(\mathsf{A}\cdot\mathsf{B}, !, \mathsf{i}, \mathsf{m})$ (resp. $(\mathsf{A}\cdot\mathsf{B}, ?, \mathsf{i}, \mathsf{m})$) shortens as $\mathsf{A}\cdot\mathsf{B}!^{\mathsf{i}}\mathsf{m}$ (resp. $\mathsf{A}\cdot\mathsf{B}?^{\mathsf{i}}\mathsf{m}$). The *subject* of an output (resp. input) event is its sender (resp. receiver); we write $\mathsf{sbj}(e)$ for the subject of event $e \in \mathcal{E}$.

The semantics of interactions, parallel composition, and iteration are straightforward. If $\mathsf{G} = \mathsf{A} \xrightarrow{\mathsf{i}} \mathsf{B} \colon \mathsf{m}$, $\leq_{\mathsf{G}}$ just establishes that the output event $\mathsf{A}\cdot\mathsf{B}!^{\mathsf{i}}\mathsf{m}$ precedes the corresponding input event $\mathsf{A}\cdot\mathsf{B}?^{\mathsf{i}}\mathsf{m}$. The semantics of parallel composition is the union of the semantics of the components, when they are defined (note that, by Definition 1, such partial orders are disjoint). If $\mathsf{G} = \mathsf{i}.\big(*\mathsf{G}'@\mathsf{A}\big)$, $\leq_{\mathsf{G}}$ is just the semantics of its body $\leq_{\mathsf{G}'}$. The other cases are more delicate.

The semantics of $\mathsf{G};\mathsf{G}'$ is defined provided that (i) the semantics of $\mathsf{G}$ and $\mathsf{G}'$ are defined, (ii) each communication event in $\mathsf{G}'$ precedes each communication event in $\mathsf{G}'$ in the order build by taking the reflexo-transitive closure of the union of $\leq_{\mathsf{G}}$, of $\leq_{\mathsf{G}'}$, and of the dependencies between the events in $\mathsf{G}$ and those in $\mathsf{G}'$ with the same subject. Fig. 2 (borrowed from [6]) illustrates how the previous construction works. The diagrams give the orders of the sequential composition of two interactions. For instance, in the first diagram the vertical arrows respectively represent the order induced by the semantics of $\mathsf{A} \xrightarrow{\mathsf{i}} \mathsf{B} \colon \mathsf{m}$ and $\mathsf{A} \xrightarrow{\mathsf{j}} \mathsf{C} \colon \mathsf{m}'$ (where $e \to e'$ means that $e$ precedes $e'$); the horizontal (and diagonal) dashed arrow is the one added because the events have the same subject. The other diagrams illustrate similar cases where sequential composition of interactions is defined, barred the last case where none of the subjects in the left and right interactions are the same. Note that this makes the sequential composition impossible as the participants in $\mathsf{C} \xrightarrow{\mathsf{j}} \mathsf{D} \colon \mathsf{m}'$ are not aware of any of the events triggered in $\mathsf{A} \xrightarrow{\mathsf{i}} \mathsf{B} \colon \mathsf{m}$.

The semantics of choice $\mathsf{G} = \mathsf{i}.\big(\mathsf{G}_1 \text{ unless } \phi_1 + \cdots + \mathsf{G}_h \text{ unless } \phi_h\big)$ is defined if the semantics of each branch $\mathsf{G}_i$ exists and $\mathsf{G}$ is *well-branched*. Roughly[6], $\mathsf{G}$ is well-branched when $(i)$ there is a unique *active* participant $\mathsf{active}(\mathsf{G})$ that chooses which branch is going to be executed and $(ii)$ any other participant $\mathsf{B}$ is *passive*, that is either $\mathsf{B}$ is oblivious of the distributed choice or $\mathsf{B}$ determines which branch is chosen from the messages it receives. If well-branchedness holds, the semantics of $\mathsf{G}$ is obtained by taking the union of the semantics of each branch $\mathsf{G}_i$ and adding the dependencies from an event marked as $\mathsf{i}$ (resp. $\mathsf{-i}$) to each minimal (resp. maximal) event of each branch. Moreover, we require that each guard $\phi_i$ predicates only on the channels of the active participant. Note that the semantics of $\mathsf{G}_1 \text{ unless } \mathtt{ff} + \cdots + \mathsf{G}_h \text{ unless } \mathtt{ff}$ is equivalent to $\mathsf{G}_1 + \cdots + \mathsf{G}_h$. In fact, removing all guards from a r-choreography yields a simple global graph: r-choreographies are a conservative extension of the global graphs in [6].

## 3 Reversible CFSMs

We adapt the definition of CFSMs in [2] to our context. A first difference is that we assume two distinguished messages $\dagger, \ddagger \in \mathcal{M}$ to respectively mark the start and the end of loops. Such messages are not explicitly used in choreographies; rather they will appear in the machines obtained by projecting global graphs in order to handle loops. Hereafter, we fix a choreography $\mathsf{G} \in \mathcal{G}$, let $\mathbf{1}$ to denote a distinguished $(unit)$ symbol, and defined the following auxiliary sets:

$$\mathtt{L}_Q = Q \times \mathcal{E} \times \varPhi \quad \text{and} \quad \overline{\mathtt{L}_Q} = \{\bar{l} \mid l \in \mathtt{L}_Q\}$$

which will respectively decorate the transitions of "ongoing" branches and those of "committed" branches. We let $\lambda$ to range over $\{\mathbf{1}\} \cup \mathtt{L}_Q \cup \overline{\mathtt{L}_Q}$ and, when $\lambda \neq \mathbf{1}$ we denote its components as $\lambda_Q$, $\lambda_{\mathcal{E}}$, and $\lambda_{\varPhi}$ (namely, $\lambda_Q = q$, $\lambda_{\mathcal{E}} = e$, and $\lambda_{\varPhi} = \phi$ when $\lambda = (q, e, \phi)$ or $\lambda = \overline{(q, e, \phi)}$).

**Definition 2 (Reversible CFSM).** *A* reversible CFSM (r-CFSM *for short*) for a participant $\mathsf{A} \in \mathcal{P}$ *is a finite transition system* $M = (Q, q_0, \rightarrow)$ *where*

- $Q$ *is a finite set of* states *with* $q_0 \in Q$ *the* initial *state, and*
- $\rightarrow \subseteq\ Q \times \mathcal{E} \times (\{\mathbf{1}\} \cup \mathtt{L}_Q \cup \overline{\mathtt{L}_Q}) \times Q$ *is a set of* transitions
- *for all transitions* $(q, e, \lambda, q') \in\rightarrow$*, we require that* $\mathit{sbj}(e) = \mathsf{A}$ *and that, if* $\lambda \neq \mathbf{1}$ *then* $\lambda_{\varPhi}$ *predicates on the channels of* $\mathsf{A}$*.*

*A* (communicating) system *is a map* $S$ *assigning a CFSM to each participant* $\mathsf{A} \in \mathcal{P}$*. Hereafter, we consider only systems consisting of deterministic CFSMs.*

Given $M = (Q, q_0, \rightarrow)$, simply write $q \xrightarrow{e} q'$ for $(q, e, \mathbf{1}, q') \in\rightarrow$ and $q \xrightarrow[\lambda]{e} q'$ for $(q, e, \lambda, q') \in\rightarrow$. We fix a system $S : \mathsf{A} \mapsto M_{\mathsf{A}}$, written $S = (M_{\mathsf{A}})_{\mathsf{A} \in \mathcal{P}}$, where $M_{\mathsf{A}}$ is the CFSM of $\mathsf{A} \in \mathcal{P}$ and without loss of generality we will assume that the local states of the machines of $S$ are pairwise disjoint; let $\mathcal{Q}_S$ be the union of the set of states of the machines $M_{\mathsf{A}}$ for $\mathsf{A} \in \operatorname{dom} S$.

[6] We refer to [6] for the formalisation of well-branchedness, which requires some technicalities that would not fit in the page limit.

As in [2], for each two participants $\mathsf{A} \neq \mathsf{B} \in \mathcal{P}$, the machines $S(\mathsf{A})$ and $S(\mathsf{B})$ communicate through two channels $\mathsf{A}{\cdot}\mathsf{B}$ and $\mathsf{B}{\cdot}\mathsf{A}$ respectively from $\mathsf{A}$ to $\mathsf{B}$ and from $\mathsf{B}$ to $\mathsf{A}$; channels have infinite capacity and behave as FIFO queues. The presence of "decorations" on transitions is the only relevant difference between Definition 2 and the standard definition of communicating machines [2]. Indeed, one can *lift* a standard CFSM $M$ to a reversible one by adding control points to each communication label (the ones above $\rightarrow$ and possibly adding triplets taken from the set $Q \times \mathcal{E} \times \Phi$ to some transitions of $M$ (the decoration below $\rightarrow$). Analogously, we can obtain a standard CFSM from a reversible one by simply "forgetting" the decorations.

Another difference wrt [2], is that we decorate messages in transit on channels with more information, in order to reverse computations. In fact, our queues store *logs*, namely tuples in the set $\mathsf{Log} \triangleq \mathcal{M} \times \mathcal{Q}_S \times \mathbb{Z}_0 \times \mathbb{N}$ (ranged over by $\ell$) where $\mathbb{N}$ is the set of natural numbers. A log $(\mathsf{m}, q, \mathsf{i}, t) \in \mathsf{Log}$, called $\mathsf{i}$-log, represents that

- either a participant, say $\mathsf{A}$, sent a message $\mathsf{m}$ explicitly specified in the choreography (namely $\mathsf{m} \neq \dagger$ and $\mathsf{m} \neq \ddagger$), or $\mathsf{A}$ entered a loop when $\mathsf{m} = \dagger$, or $\mathsf{A}$ exited a loop when $\mathsf{m} = \ddagger$,
- state $q$ is the state the sender $\mathsf{A}$ was in when $\mathsf{m}$ was sent,
- $\mathsf{i}$ is the control point of the interaction of $\mathsf{G}$ where $\mathsf{m}$ originated, and
- "time-stamp" $t$ represents the fact that $\mathsf{m}$ was the $t$-th message sent by $\mathsf{A}$.

In fact, our channels will be associated with elements of $\mathsf{Log}^\star \times \mathsf{Log}^\star$ that we write as

$$\ell_1, \cdots, \ell_{p-1} \ ; \ \ell_p, \cdots, \ell_h \tag{7}$$

to emphasise that the sequence $\ell_1, \cdots, \ell_{p-1}$ in (7) yields the messages already consumed by the receiver, while the sequence $\ell_p, \cdots, \ell_h$ yields the pending message yet to be consumed. We require[7] that for each $1 \leq i < j \leq h$ if $\ell_i = (_, _, _, t)$ and $\ell_j = (_, _, _, t')$ then $t < t'$, in which case we say that $\ell_i$ is *before* $\ell_j$ (or $\ell_j$ is *after* $\ell_i$) in the channel (7). Basically, (7) can be thought of as a pair of two totally ordered sets of logs where the order is established by the time-stamps.

## 4 Forward Semantics

Following [2], our semantics describes how a *configuration* evolves due to the communication actions of the machines. Essentially, a *configuration* of $S$ keeps track of the state of each machine and the content of each channel in $S$, however we need to enrich configurations wrt [2] to account for reversibility.

[7] We use $_$ as a wild-card where universally quantified meta-variables are expected.

$$[\textbf{out}]\quad \frac{\sigma(\mathsf{A}) \xrightarrow[\lambda]{\mathsf{A}\cdot\mathsf{B}!^{\mathsf{i}}\mathsf{m}} q \qquad \ell = (\mathsf{m}, \sigma(\mathsf{A}), \mathsf{i}, t) \qquad t = 1 + \max_{\mathsf{C}\in\mathcal{P}\setminus\{\mathsf{A}\}} \{t' | (_, _, _, t') \in \chi(\mathsf{A}\cdot\mathsf{C})\}}{\langle \sigma, \chi, \beta\rangle \Longrightarrow \langle \sigma[\mathsf{A} \mapsto q], \chi[\mathsf{A}\cdot\mathsf{B} \mapsto \chi(\mathsf{A}\cdot\mathsf{B}).\ell], \mathtt{upd_O}(\lambda, \beta)\rangle}$$

$$[\textbf{inp}]\quad \frac{\sigma(\mathsf{A}) \xrightarrow[\lambda]{\mathsf{B}\cdot\mathsf{A}?^{\mathsf{i}}\mathsf{m}} q \qquad \chi(\mathsf{B}\cdot\mathsf{A}) = \ell_1, \cdots, \ell_{p-1} \ ; \ \ell_p, \cdots, \ell_h \qquad \ell_p = (\mathsf{m}, _, _, _)}{\langle \sigma, \chi, \beta\rangle \Longrightarrow \langle \sigma[\mathsf{A} \mapsto q], \chi[\mathsf{A}\cdot\mathsf{B} \mapsto \ell_1, \cdots, \ell_{p-1}, \ell_p \ ; \ \ell_{p+1}, \cdots, \ell_h], \mathtt{upd_I}(\lambda, \beta)\rangle}$$

**Fig. 3.** Rules for forward semantics

**Definition 3 (Configuration & history).** *A* configuration *of $S$ is a tuple $\langle \sigma, \chi, \beta\rangle$ where $\sigma : \mathcal{P} \to \mathcal{Q}_S$ (called* machines configuration*) is such that, for each $\mathsf{A} \in \mathcal{P}$, $\sigma(\mathsf{A}) \in Q_{\mathsf{A}}$, $\chi : \mathcal{C} \to \mathsf{Log}^\star \times \mathsf{Log}^\star$ (called* channels configuration*), and $\beta : \mathcal{Q}_S \to 2^{\mathcal{E}\times\Phi} \times \{\mathtt{tt}, \mathtt{ff}\}$. We write $\mathcal{H}(\chi)$ for the* history *of $\chi$, that is the map returning the first component of $\chi(\mathsf{A}\cdot\mathsf{B})$ for each channel $\mathsf{A}\cdot\mathsf{B} \in \mathcal{C}$.*

A configuration $\langle \sigma, \chi, \beta\rangle$ is a snapshot of $S$ where the current *local* state of each machine $\mathsf{A}$ is given by $\sigma(\mathsf{A})$, the state of the channels is given by $\chi$, and $\beta$ records which branches have been tried and a boolean which becomes true when backward computations are no longer meaningful (because all branches have been already tried or all have their guards true). Also, we write $\ell \in \chi$ when there is a channel $\mathsf{A}\cdot\mathsf{B} \in \mathcal{C}$ such that $\ell$ is an element of either one of the components of $\chi(\mathsf{A}\cdot\mathsf{B})$ and similarly $\ell \in \mathcal{H}(\chi)$ means that there is a channel $\mathsf{A}\cdot\mathsf{B} \in \mathcal{C}$ such that $\ell$ is an element of $\mathcal{H}(\chi(\mathsf{A}\cdot\mathsf{B}))$.

The forward semantics of $S$ (cf. Definition 4 below) is defined in terms of configurations reachable from the *initial configuration of* $S$, that is from configuration $\langle \sigma_0, \chi_0, \beta_0\rangle$ where $\sigma_0 : \mathsf{A} \mapsto q_{0\mathsf{A}}$ is the map taking each participant $\mathsf{A} \in \mathcal{P}$ to its initial state $q_{0\mathsf{A}}$, $\chi_0 : \mathsf{A}\cdot\mathsf{B} \mapsto \varepsilon \ ; \ \varepsilon$, where $\varepsilon \in \mathsf{Log}^\star$ is the empty sequence of logs, and $\beta_0 : q \mapsto (\emptyset, \mathtt{ff})$ for each $q \in \mathcal{Q}_S$. A decoration $\lambda \in \{\mathbf{1}\} \cup \mathrm{L}_Q \cup \overline{\mathrm{L}_Q}$ is *valid* for $\beta$ when $\lambda = \mathbf{1}$ or $\big((\lambda_{\mathcal{E}}, \lambda_\Phi) \in \beta(\lambda_Q) \implies \beta(\lambda_Q) = (_, \mathtt{tt})\big)$. Before formally defining the forward semantics we need to put in place two auxiliary functions $\mathtt{upd_O} : (\lambda, \beta) \mapsto \beta'$ and $\mathtt{upd_I} : (\lambda, \beta) \mapsto \beta'$ which update the mapping $\beta$ according to the decoration $\lambda$:

$$\mathtt{upd_O}(\lambda, \beta) = \begin{cases} \beta & \lambda \notin \overline{\mathrm{L}_Q} \text{ and if } \lambda \in \mathrm{L}_Q \text{ then } \lambda \text{ valid for } \beta \\ \\ \beta[\lambda_Q \mapsto (\emptyset, \mathtt{ff})] & \lambda \in \overline{\mathrm{L}_Q} \text{ and } \lambda \text{ valid for } \beta \end{cases}$$

$$\mathtt{upd_I}(\lambda, \beta) = \begin{cases} \beta & \lambda \notin \overline{\mathrm{L}_Q} \\ \\ \beta[\lambda_Q \mapsto (\emptyset, \mathtt{ff})] & \lambda \in \overline{\mathrm{L}_Q} \end{cases}$$

(note that $\mathtt{upd_O}$ is a partial map). If $\lambda$ is the decoration of a transition not in the branch or of one in an ongoing branch and its guard still holds then nothing changes in $\beta$. If $\lambda$ decorates a transition leaving a branch whose guard still holds, then $\beta$ is reset to the empty set. Likewise for $\mathtt{upd_I}$.

**Definition 4 (Forward semantics).** *Let* $\Longrightarrow$ *be the smallest relation between configurations closed under the rules of Fig. 3 with* [OUT] *applicable only if* $\mathtt{upd_0}(\lambda, \beta)$ *is defined. If a configuration* $\langle \sigma, \chi, \beta \rangle$ *and a configuration* $\langle \sigma', \chi', \beta' \rangle$ *are related by reflexo-transitive of* $\Longrightarrow$ *then we say that the former* reaches *the second. The set of* (forward) reachable configurations of $S$ *is* $\mathsf{RC}_S \triangleq \{\langle \sigma, \chi, \beta \rangle \mid \langle \sigma_0, \chi_0, \beta_0 \rangle$ *reaches* $\langle \sigma, \chi, \beta \rangle\}$*.*

The rules in Fig. 3 specify how machines produce (rule [OUT]) or consume (rule [IMP]) messages. The semantics of the output of a message differs depending on whether the action is ($\lambda \neq \mathbf{1}$) on a brach of a choice or not ($\lambda = \mathbf{1}$)

The second case basically reproduces the standard behaviour of CFSMs. In fact, the rule simply enqueues a log carrying the information about the actual message m, the state $\sigma(\mathsf{A})$ from which A sends m, the corresponding control point i, and the time-stamp $t$ computed as the increment of the maximal time-stamp on the channels from A. Besides updating the channel A·B, in the reached configuration we record that the new local state of A is $q$. An output on a branching behaves exactly as a "normal" output except that, when the branch can still be reverted ($\lambda$ is valid in $\beta$), the guard of the branch does not hold. Moreover, if the transition is a committed branch, say $\lambda \in \overline{\mathtt{L}_Q}$, then the information about the branches of that choice are reset. In this way the mapping corresponding to the (entering choice) state $\lambda_Q$ is set to $\emptyset$. As we will see, the backward semantics will set $\beta(\lambda_Q) = (_, \mathtt{tt})$ when all branches have been tried, enabling the forward semantics to take any of the branches when no other option is possible. The semantics of input actions is very simple: the machine consumes the message on the top of the queue if it matches the one in the transition from the local state. Moreover, if the input is annotated with a brach exiting decoration ($\lambda \in \overline{\mathtt{L}_Q}$), then the mapping $\beta$ is updated accordingly, otherwise $\beta$ is left unchanged. Notice that the consumption of the message is represented by the fact that the corresponding log "becomes history".

## 5 Projections of r-CFSMs

We will consider systems consisting of machines obtained by determinising the automata "projected" from an r-choreography. The projection of a global graph, defined in the rest of this section, extends the operation defined in [6] to account for guards. The construction is very similar to the one in [6] and simply requires to decorate the transitions on branches.

It is convenient to consider *pseudo-machines* (shortened p-machine), namely r-CFSMs with an "interface" consisting of a special state $q_e$; when composing p-machines, transitions to/from other machines may start from/land on $q_e$. Call $\mathtt{M} = (Q\ ; q_0\ ; q_e\ ; \rightarrow)$ a *p-machine* when $(Q, q_0, \rightarrow)$ is an r-CFSM and $q_e \in Q$. We define

$$\mathtt{M} \ \%\ (\hat{q}, \hat{e}, \phi) = (Q \cup \{\hat{q}\}\ ; q_0\ ; q_e\ ; \rightarrow')$$

where

$$\rightarrow' = \{(q, (e, (\hat{q}, \hat{e}, \phi)), q') \mid (q, e, q') \in \rightarrow \wedge\, q' \neq q_e\} \cup$$
$$\{(q, (e, \overline{(\hat{q}, \hat{e}, \phi)}), q') \mid (q, e, q') \in \rightarrow \wedge\, q' = q_e\}$$

Also, given two p-machine $\mathtt{M}_i = (Q_i\ ; q_{0i}\ ; q_{ei}\ ; \rightarrow_i)$ with $i \in \{1, 2\}$, we write $\mathtt{M}_1 \cap \mathtt{M}_2$ for $Q_1 \cap Q_2$. We define the following operations when $\mathtt{M}_1 \cap \mathtt{M}_2 = \emptyset$:

$$\mathtt{M}_1 ; \mathtt{M}_2 = ((Q_1 \cup Q_2)\theta\ ; q_{01}\ ; q_{e2}\ ; (\rightarrow_1 \cup \rightarrow_2)\theta) \text{ with } \theta = [q_{e1} \mapsto q_{02}]$$
$$\mathtt{M}_1 \sqcup \mathtt{M}_2 = ((Q_1 \cup Q_2)\theta\ ; q_{01}\ ; q_{e2}\ ; (\rightarrow_1 \cup \rightarrow_2)\theta) \text{ with } \theta = [q_{e1}, q_{01} \mapsto q_{e2}, q_{02}]$$

where $\theta = [q_1 \mapsto q_2]$ (read $q_1$ is replaced with $q_2$) is the substitution operation on sets, transition relations and decorations defined as follows:

$$Q[q_1 \mapsto q_2] = Q \setminus \{q_1\} \cup \{q_2\} \text{ if } q_1 \in Q \qquad Q[q_1 \mapsto q_2] = Q \text{ if } q_1 \notin Q$$

$$\rightarrow[q_1 \mapsto q_2] = \bigcup_{(q,\lambda,q') \in \rightarrow} (q[q_1 \mapsto q_2], \lambda[q_1 \mapsto q_2], q'[q_1 \mapsto q_2])$$

$$(q, e, \phi)[q_1 \mapsto q_2] = (q[q_1 \mapsto q_2], e, \phi) \qquad \overline{(q, e, \phi)}[q_1 \mapsto q_2] = \overline{(q[q_1 \mapsto q_2], e, \phi)}$$

$$q_1[q_1 \mapsto q_2] = q_2 \qquad q[q_1 \mapsto q_2] = q \text{ if } q \neq q_1 \qquad \mathbf{1}[q_1 \mapsto q_2] = \mathbf{1}$$

Given two substitutions $\theta_1 = [q_1 \mapsto q_2]$ and $\theta_2 = [q_3 \mapsto q_4]$, we indicate with $\theta_1, \theta_2$ the concatenation of two substitutions. Sometimes we will write $[q_1, q_3 \mapsto q_2, q_4]$ instead of $\theta_1, \theta_2$. Moreover we will write $q[q_1 \mapsto q_2]$ instead of $\{q\}[q_1 \mapsto q_2]$.

We can now show how to project a global graph. To this purpose, the map $\mathbf{proj}_{\mathsf{A}}(\mathsf{G}, _, _)$, given a global graph $\mathsf{G} \in \mathcal{G}$, returns the p-machine wrt participant $\mathsf{A} \in \mathcal{P}$; the last two parameters are the initial state of the resulting machine and its interface state. We define $\mathbf{proj}_{\mathsf{A}}(\mathsf{G}, _, _)$ by induction on the structure of $\mathsf{G}$. The base case is as follows[8]:

$\mathbf{proj}_{\mathsf{A}}(\mathsf{G}, q_0, q_e) =$

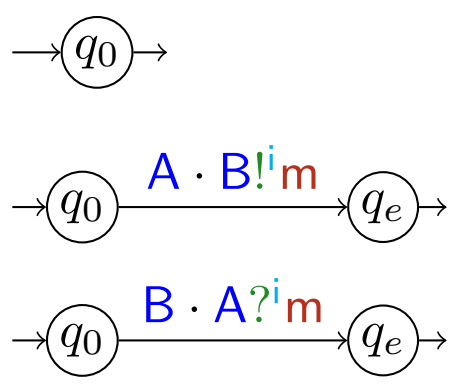

if $\mathsf{G} = \mathsf{B} \xrightarrow{\mathsf{i}} \mathsf{C} \colon \mathsf{m}$

if $\mathsf{G} = \mathsf{A} \xrightarrow{\mathsf{i}} \mathsf{B} \colon \mathsf{m}$

if $\mathsf{G} = \mathsf{B} \xrightarrow{\mathsf{i}} \mathsf{A} \colon \mathsf{m}$

Basically, the projection of A for interaction yields a p-machine without transitions if the participants of the interaction are different from A or the p-machine with the output or the input transition corresponding to the interaction depending on whether A is the sender or the receiver of the interaction.

[8] We use a graphical notation for p-machine where the initial state is marked by a dandling arrow with no source and the interface is marked by a dandling arrow with no target.

For sequential composition we have:

$$\mathbf{proj}_{\mathsf{A}}(\mathsf{G};\mathsf{G}', q_0, q_e) = \mathtt{M};\mathtt{M}' \text{ where } \mathtt{M} = \mathbf{proj}_{\mathsf{A}}(\mathsf{G}, q_0, {q_e}'), \mathtt{M}' = \mathbf{proj}_{\mathsf{A}}(\mathsf{G}', {q_e}', q_e),$$
$$\text{and } \mathtt{M} \cap \mathtt{M}' = \{{q_e}'\}$$

which is straightforward; note that the projection is defined only if the p-machine share only ${q_e}'$.

For the choice we have:

$$\mathbf{proj}_{\mathsf{A}}(\mathsf{i}.(\mathsf{G}_1 \ {}_{\mathtt{unless}}\ \phi_1 + \cdots + \mathsf{G}_h \ {}_{\mathtt{unless}}\ \phi_h), q_0, q_e) = \mathtt{M}_1 \sqcup \cdots \sqcup \mathtt{M}_h$$
$$\text{where } \forall 1 \le j \le h : \ \mathtt{M}_j = \begin{cases} (\mathbf{proj}_{\mathsf{A}}(\mathsf{G}_j, q_0, q_e))\%(q_0, e_j, \phi_j), & \text{if } \mathsf{A} = \mathsf{active}(\mathsf{G}) \\ \mathbf{proj}_{\mathsf{A}}(\mathsf{G}_j, q_0, q_e), & \text{otw} \end{cases}$$

and we assume $\mathtt{M}_j \cap \mathtt{M}_{j'} = \{q_0, q_e\}$ for all $1 \le j \ne j' \le h$ and $e_j$ is the first output of $\mathsf{A}$ on the branch $1 \le j \le h$. The projection of a choice consists of the projections of each branches departing from the initial state $q_0$; moreover, the active transitions of the active participant $\mathsf{A}$ are decorated in order to decide when to reverse the computation.

The projection of a parallel composition is just the automaton product of the projections of the threads:

$$\mathbf{proj}_{\mathsf{A}}(\mathsf{G} \mid \mathsf{G}', q_0, q_e) = \mathbf{proj}_{\mathsf{A}}(\mathsf{G}, \overline{q_0}, \overline{q_e}) \times \mathbf{proj}_{\mathsf{A}}(\mathsf{G}', \underline{q_0}, \underline{q_e})$$

where $q_0 = (\overline{q_0}, \underline{q_0})$, $q_e = (\overline{q_e}, \underline{q_e})$, and $\mathbf{proj}_{\mathsf{A}}(\mathsf{G}, \overline{q_0}, \overline{q_e}) \cap \mathbf{proj}_{\mathsf{A}}(\mathsf{G}', \underline{q_0}, \underline{q_e}) = \emptyset$.

Finally, the projection of an iterated choreography is as follows:

$$\mathbf{proj}_{\mathsf{A}}(\mathsf{i}.(*\mathsf{G}@\mathsf{C}), q_0, q_e) = \mathtt{M}_{\text{entry}}; (\mathtt{M}_{\text{body}} \sqcup \mathtt{M}_{\text{exit}})$$

where $\mathtt{M}_{\text{body}} = \mathbf{proj}_{\mathsf{A}}(\mathsf{G}, \underline{q_0}, \underline{q_e})$ while $\mathtt{M}_{\text{entry}}$ and $\mathtt{M}_{\text{exit}}$ depend on whether $\mathsf{A} = \mathsf{C}$ or not; and we have

if $\mathsf{A} = \mathsf{C}$ then

$\mathtt{M}_{\text{entry}} = \prod_{\mathsf{B} \in \mathsf{G} \setminus \{\mathsf{A}\}}$ 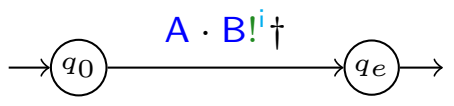

$\mathtt{M}_{\text{exit}} = \prod_{\mathsf{B} \in \mathsf{G} \setminus \{\mathsf{A}\}}$ 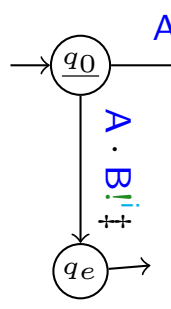

if $\mathsf{A} \ne \mathsf{C}$ then

$\mathtt{M}_{\text{entry}} =$ 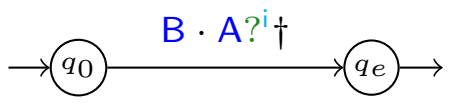

$\mathtt{M}_{\text{exit}} =$ 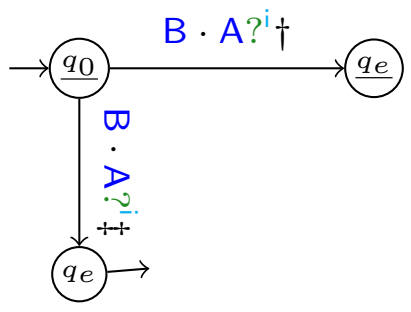

Basically, the projection of the active participant requires to send all the other participants in the body the start loop message † before "entering" the body and the start or end loop message ‡ after the body depending on if $\mathsf{A}$ wants to repeat or leave the loop. Such outputs can be sent in any order, and in fact we consider all the interleavings taking the product of the machines. For the other participants, we just have to receive the start message and wait for the decision on whether repeat or leave the loop.

Finally, the projected machine for a participant $\mathsf{A}$ is the r-CFSM obtained by removing the interface from $\mathbf{proj}_{\mathsf{A}}(\mathsf{G}, q_0, q_e)$ and minimising the resulting automaton wrt language equivalence.

*Example 3.* The r-CFSMs obtained projecting the r-choreography in Example 1 are reported in Appendix A.

## 6 Backward Semantics

A key design choice of our framework is that we opt for a non fully reversible model, that is for a model where not all computations are reversible. For us, reversing computations boils down to (possibly) undo branches when their guard becomes true. To this purpose we appeal to a relation $\chi \vdash \phi$ that determines when a configuration entails a guard. Roughly, outputs are reversible if none of their effects has been consumed by an input. The main motivation of this choice is that it would be hard to coordinate the undo of an input since the sender does not block; we discuss other options in Section 7.

Nonetheless, we can possibly reverse inputs when they are within a loop. In fact, an interesting aspect of our approach is how we deal with iterative choreographies. When is a loop reversible? A possible answer to such question is to reverse the loop provided that it is still "ongoing". We will make this concept precise; for the moment, one can think that a loop is ongoing until none of the involved participants has consumed the end loop message. This choice is motivated by the fact that we consider choreographies as a specification language where one abstracts away by the actual termination conditions of loops; hence, until a loop is not exited, we can safely undo its iterations. This allows to undo computations where some messages sent in a loop have been consumed by the receiver. Note that inputs outside loops are irreversible.

**Definition 5 (Loops).** *The set* $\mathtt{loop}(\mathsf{G})$ *of* loops of $\mathsf{G}$ *is inductively defined as*

$$\mathtt{loop}(\mathsf{A} \xrightarrow{i} \mathsf{B} \colon \mathsf{m}) = \emptyset \qquad \mathtt{loop}(\mathsf{G};\mathsf{G}') = \mathtt{loop}(i.(\mathsf{G} \mid \mathsf{G}')) = \mathtt{loop}(\mathsf{G}) \cup \mathtt{loop}(\mathsf{G}')$$

$$\mathtt{loop}(i.(*\mathsf{G}@\mathsf{A})) = \{i.(*\mathsf{G}@\mathsf{A})\} \cup \mathtt{loop}(\mathsf{G})$$

$$\mathtt{loop}(i.(\mathsf{G}_1 \ {\scriptstyle\mathtt{unless}}\ \phi_1 + \cdots + \mathsf{G}_h \ {\scriptstyle\mathtt{unless}}\ \phi_h)) = \bigcup_{i \in 1,\ldots,h} \mathtt{loop}(\mathsf{G}_i)$$

*An* $i$*-log* $\ell \in \mathsf{Log}$ belongs to a loop $\mathsf{L} \in \mathtt{loop}(\mathsf{G})$*, in symbols* $\ell \in \mathsf{L}$*, when* $i \in cp(\mathsf{L})$*. Given* $\mathsf{L} = i.(*\mathsf{G}@\mathsf{A})$*,* $i$ *is the* loop id *of* $\mathsf{L}$*, a log of the form* $\ell = (\dagger, q, i, t)$ *is a* start loop log *of* $\mathsf{L}$*, and a log of the form* $\ell = (\ddagger, q, i, t)$ *is an* end loop log *of* $L$*.*

A key concept to define the backward semantics of r-CFSMs is *run-time causality.* This notion relies on the semantic relations $\leq_{\mathsf{G}}$ on the events of $\mathsf{G}$ (and it is therefore not defined when $\leq_{\mathsf{G}}$ is undefined). Intuitively, this relation established the order of arrival of logs in the channels. Such order allows us to identify those logs that can be reverted and the order in which reverse the communications. Crucially loops complicate the definition of run-time causality since when the interaction $\mathsf{A} \xrightarrow{\mathsf{i}} \mathsf{B}\colon \mathsf{m}$ is on loop there can possibly be many instances of $\mathsf{i}$-logs in the channel $\mathsf{A}{\cdot}\mathsf{B}$. Note that we only allow well-structured nested loops in our r-choreographies, therefore if $\ell \in \mathsf{L}$ and $\ell \in \mathsf{L}'$ then either $\mathsf{L}$ is a sub-term of $\mathsf{L}'$ or vice versa. If $\mathsf{i} \in \mathsf{cp}(\mathsf{G})$ then $\mathsf{event}(\mathsf{i}) = \mathsf{A} \cdot \mathsf{B}!^{\mathsf{i}}\mathsf{m}$ if $\mathsf{i}$ is the control point of an interaction $\mathsf{A} \xrightarrow{\mathsf{i}} \mathsf{B}\colon \mathsf{m}$ and $\mathsf{event}(\mathsf{i}) = \mathsf{i}$ otherwise. In the following definitions we assume fixed a channels configuration $\chi : \mathcal{C} \to \mathsf{Log}^\star \times \mathsf{Log}^\star$ and, for a log $\ell \in \chi(\mathsf{A}{\cdot}\mathsf{B})$ on a channel $\mathsf{A}{\cdot}\mathsf{B}$, we say that $\ell$ *is on the the* $(n+1)$*-th iteration of* $\mathsf{L}$ *in* $\chi$ (with $n \geq 0$) if $\ell \in \mathsf{L}$, $\ell$ is before the $(n+1)$-th end loop log of $\mathsf{L}$ in $\chi(\mathsf{A}{\cdot}\mathsf{B})$ and $\ell$ is after the $n$-th end loop log when $n > 0$ and it is after the first start loop log when $n = 0$.

**Definition 6 (Runtime causality).** *The relation $\sqsubset_\chi$ on logs of $\chi$ is the minimal relation closed under the following rules:*

1. *if there is $\mathsf{A} \cdot \mathsf{B} \in \mathcal{C}$ such that $\ell, \ell' \in \chi(\mathsf{A}{\cdot}\mathsf{B})$ with $\ell = (_,_,_,t)$ and $\ell' = (_,_,_,t')$ then $\ell \sqsubset_\chi \ell'$ if $t < t'$*
2. *if there are $\mathsf{A} \cdot \mathsf{B} \neq \mathsf{A}' \cdot \mathsf{B}' \in \mathcal{C}$ such that $(\mathsf{m}, q, i, t) = \ell \in \chi(\mathsf{A}{\cdot}\mathsf{B})$ and $(\mathsf{m}', q', j, t') = \ell' \in \chi(\mathsf{A}'{\cdot}\mathsf{B}')$ and $\mathsf{event}(i) \leq_{\mathsf{G}} \mathsf{event}(j)$ then*
   - *(a) if $\forall \mathsf{L} \in \mathtt{loop}(\mathsf{G}) : \ell, \ell' \notin \mathsf{L} \vee (\ell \in \mathsf{L} \iff \ell' \notin \mathsf{L})$ then $\ell \sqsubset_\chi \ell'$*
   - *(b) otherwise, let $\mathsf{L} \in \mathtt{loop}(\mathsf{G})$ be the smallest loop such that $\ell, \ell' \in \mathsf{L}$, let $\ell$ and $\ell'$ respectively be on the $(n+1)$-th and $(m+1)$-th rounds of $\mathsf{L}$ in their channels, then*
     - *if $n \leq m$ then $\ell \sqsubset_\chi \ell'$*
     - *if $n > m$ then $\ell' \sqsubset_\chi \ell$*

*Moreover we define the relation $\sqsubseteq_{\mathsf{G}}$ as follows: $\ell \sqsubseteq_{\mathsf{G}} \ell' \iff \ell = \ell' \vee \ell \sqsubset^+_{\mathsf{G}} \ell'$, where $\sqsubset^+_{\mathsf{G}}$ is the transitive closure of $\sqsubset_{\mathsf{G}}$.*

By condition 1 in Definition 6, the causal relation between logs on a same channels is established by their "temporal" ordering. The most difficult part is to establish a causality relation between logs on different channels. For this, condition 2 requires that the events $\mathsf{event}(\mathsf{i})$ and $\mathsf{event}(\mathsf{j})$ corresponding to $\ell$ and $\ell'$ respectively are events ordered by the semantics of choreographies. Moreover, it is necessary to analyse the loop structure involving $\ell$ and $\ell'$. If the logs are not an a loop or they are on disjoint loops, then they are ordered at run-time because each $\mathsf{j}$-log has to follow each of $\mathsf{i}$-log. When there is a loop $\mathsf{L}$ including both $\ell$ and $\ell'$ the situation is more delicate because some $\mathsf{i}$-logs will causally depend at run-time by some $\mathsf{j}$-logs even if $\mathsf{event}(\mathsf{i}) \leq_{\mathsf{G}} \mathsf{event}(\mathsf{j})$ due to the "unfolding" of $\mathsf{L}$. In this case we check the rounds of $\mathsf{L}$ on the different channels. When $\ell$ and $\ell'$ are on the same round of $\mathsf{L}$ or the former is on an earlier round than the latter, then any $\mathsf{i}$-log in the round precedes any $\mathsf{j}$-log in the same round. Instead, $\ell'$ precedes $\ell$ when the former is on an earlier round than the latter.

We can now define the points of a configuration where roll-backs are allowed. For this we need to decide when loops are not completed yet. We say that $\mathsf{L}$ is *ongoing* if there is $\mathsf{A}\cdot\mathsf{B} \in \mathcal{C}$ such that there is a start loop log of $\mathsf{L}$, say $\ell \in \chi(\mathsf{A}\cdot\mathsf{B})$, and either all end loop logs of $\mathsf{L}$ are before $\ell$ in $\chi(\mathsf{A}\cdot\mathsf{B})$ or there is $\ell' \in \chi(\mathsf{A}\cdot\mathsf{B}) \setminus \mathcal{H}(\chi(\mathsf{A}\cdot\mathsf{B}))$ end loop log of $\mathsf{L}$, where given the content of a channel $\ell_1, \cdots, \ell_{p-1}\ ;\ \ell_p, \cdots, \ell_h \in \mathsf{Log}^\star \times \mathsf{Log}^\star$ and a set of logs $X \subseteq \mathsf{Log}$, we define $(\ell_1, \cdots, \ell_{p-1}\ ;\ \ell_p, \cdots, \ell_h) \setminus X \triangleq (\ell_1, \cdots, \ell_{p-1}) \setminus X\ ;\ (\ell_p, \cdots, \ell_h) \setminus X$ as the content where the log in $X$ are removed from the channel.

Hence, a log can be reversed when all its effects are not in the history of the configuration, unless they are on an ongoing loop as required by Definition 7 below.

**Definition 7 (Roll-back point).** *Let* $\mathsf{AB} \in \mathcal{C}$ *be a channel. The set* $\mathtt{RBP}_{\mathsf{G},\chi}(\mathsf{A}\cdot\mathsf{B})$ *of* roll-back points in $\chi$ for a channel $\mathsf{A}\cdot\mathsf{B}$ (wrt $\mathsf{G}$) *is the set of logs* $\ell \in \chi(\mathsf{A}\cdot\mathsf{B})$ *such that*

- *if* $\ell$ *is not on a loop then, for all* $\ell' \in \chi$, $\ell \sqsubseteq_G \ell'$ *implies* $\ell' \notin \mathcal{H}(\chi)$
- *if there is* $\mathsf{L} \in \mathtt{loop}(\mathsf{G})$ *s.t.* $\ell \in \mathsf{L}$ *then, for all* $\ell' \in \mathcal{H}(\chi)$ *and taken* $\mathsf{L}'$ *to be the outermost loop s.t.* $\mathit{cp}(\mathsf{L}) \subseteq \mathit{cp}(\mathsf{L}')$*, we have that* $\ell \sqsubseteq_G \ell'$ *implies that* $\ell' \in \mathsf{L}'$ *and* $\mathsf{L}'$ *is ongoing.*

To reverse a communication we define an operation that "undoes" reversible positions starting from the maximal ones until there are no more positions to reverse. Given a machines configuration $\sigma : \mathcal{P} \to \mathcal{Q}_S$ and a set of logs $T \subseteq \mathsf{Log}$, we define:

$$\rho(T, \sigma, \chi) = \begin{cases} \langle \sigma, \chi \rangle & \text{if } T = \emptyset \\ \rho(T \setminus \{\ell\}, \sigma[\mathsf{S} \mapsto q], \chi \setminus \{\ell\}) & \text{otw } \ell \in \mathtt{max}_{\mathsf{G},\chi}(T)\ \ \mathsf{i} - \text{log},\ \ \mathsf{event}(\mathsf{i}) = \mathsf{S} \cdot \mathsf{R}!^{\mathsf{i}}\mathsf{m} \end{cases}$$

where $\mathtt{max}_{\mathsf{G},\chi}(T) = \{\ell \in T \,|\, \forall \ell' \in T : \ell \not\sqsubseteq_G \ell'\}$ and $\chi \setminus \{\ell\} : \mathsf{A}\cdot\mathsf{B} \mapsto \chi[\mathsf{A}\cdot\mathsf{B} \mapsto \chi(\mathsf{A}\cdot\mathsf{B}) \setminus \{\ell\}]$. Basically, $\rho$ starts removing the maximal logs in $T$ from the configuration while updating the state of each machine to the state from where such logs had been sent. Note that $\rho$ is well defined as its result does not depend on the order in which independent maximal elements of $T$ are removed (since they are independent).

Finally, the backward semantics is the relation specified by the following rule:

$$[\textbf{rev}]\quad \frac{\begin{array}{c} \sigma(\mathsf{A}) \xrightarrow[\lambda]{-} \qquad \lambda_{\mathcal{E}} = \mathsf{A}\cdot\mathsf{B}!^{\mathsf{i}}\mathsf{m} \\ \beta(\lambda_Q) = (_, \mathtt{ff}) \qquad \chi \vdash \lambda_\phi \qquad \ell = (\mathsf{m}, \lambda_Q, \mathsf{i}, _) \in \mathtt{RBP}_{\mathsf{G},\chi}(\mathsf{A}\cdot\mathsf{B}) \\ T = \{\ell' \in \chi \,|\, \ell \sqsubseteq_{\mathsf{G}} \ell'\} \qquad b = \forall \hat{q} \xrightarrow[\lambda']{-}\ :\ ((\lambda'_{\mathcal{E}}, \lambda'_{\Phi}) \in \beta(\hat{q}) \vee \chi \vdash \lambda'_{\Phi}) \end{array}}{\langle \sigma, \chi, \beta \rangle \rightsquigarrow \langle \rho(T, \sigma, \chi), \beta[\hat{q} \mapsto (\beta(\hat{q}) \cup \{(\mathsf{A}\cdot\mathsf{B}!^{\mathsf{i}}\mathsf{m}, \lambda_\phi)\}, b)] \rangle}$$

When participant $\mathsf{A}$ is in a state of a branch (condition $\sigma(\mathsf{A}) \xrightarrow[\lambda]{-}$) that has not been fully explored yet (condition $\beta(\lambda_Q) = (_, \mathtt{ff})$) and whose guard is satisfied (condition $\chi \vdash \lambda_\phi$), we can reverse any log which is a roll-back point in $\chi$ of the selecting message $\mathsf{m}$. Observe that rule [REV] updates the information in $\beta$ by recording that the branch corresponding to $\mathsf{A}\cdot\mathsf{B}!^{\mathsf{i}}\mathsf{m}, \lambda_\phi$ is being tried and setting the flag $b$ to $\mathtt{tt}$ if no more branches from $\hat{q}$ are left to be attempted.

*Example 4.* A backward execution of our running example are in Appendix A.

We now state some properties of our constructions. Firstly, there is a strict correspondence between standard and reversible CFSMs. As observed in Section 3, given a r-CFSM $M = (Q, q_0, \to)$, we can define the standard machine $\mathsf{forget}(M) = (Q, q_0, \to')$ by simply ignoring decorations, that is by defining:

$$\to' = \bigcup_{(q,e,\lambda,q') \in \to} (q, e, q')$$

Let $\mathsf{forget}(S) : \mathsf{A} \mapsto \mathsf{forget}(S(\mathsf{A}))$ be the system of standard CFSMs induced by the system $S$ of r-CFSMs. Also, overloading $\mathsf{forget}$ on $\mathsf{Log}$, $\mathsf{Log}^\star$, and on configurations, we can define

$$\mathsf{forget} : (\mathsf{m}, q, \mathsf{i}, t) \mapsto \mathsf{m} \quad \mathsf{forget} : \ell_1, \cdots, \ell_n \mapsto \mathsf{forget}(\ell_1), \cdots, \mathsf{forget}(\ell_n) \quad \text{and}$$
$$\mathsf{forget} : \langle \sigma, \chi, \beta \rangle \mapsto \langle \sigma, \mathsf{A}{\cdot}\mathsf{B} \mapsto \mathsf{forget}(\mathsf{forget}(\ell_1), \cdots, \mathsf{forget}(\ell_n)) \rangle \text{ if } \chi : \mathsf{A}{\cdot}\mathsf{B} \mapsto _ \,;\, \ell_1, \cdots, \ell_n$$

If we extend the operation $\%$ to sets of reversion labels, i.e. subsets of $Q \times \mathcal{E}^! \times \Phi$, then for each r-CFSMs $M$ there are a CFSM $\hat{M}$ and a set of reversion labels $X$ such that $M = \hat{M}\%X$. It is straightforward to show the following

**Fact 1.** *For each CFSM $M$ and each $X \subset Q \times \mathcal{E}^! \times \Phi$,* $\mathsf{forget}(M\%X) = M$*.*

The following two theorems prove that our forward semantics is a simple decoration of the standard semantics of CFSMs.

**Theorem 1 (Forward soundness).** *If $\langle \sigma, \chi, \beta \rangle \in \mathsf{RC}_S$ then* $\mathsf{forget}(\langle \sigma, \chi, \beta \rangle) \in \mathsf{RC}_{\mathsf{forget}(S)}$.

**Theorem 2 (Forward completeness).** *If $S'$ is a system of standard CFSMs such that $S' = \mathsf{forget}(S)$, then for all $\langle \sigma, \chi \rangle \in \mathsf{RC}_{S'}$ there is a configuration $\langle \sigma', \chi', \beta' \rangle \in \mathsf{RC}_S$ such that* $\mathsf{forget}(\langle \sigma', \chi', \beta' \rangle) = \langle \sigma, \chi \rangle$*.*

Theorem 3 below establishes *causal consistency*, a typical property of "reasonable" reversible computations. Causal consistency requires indeed that backward semantics does not introduce configurations that are not reachable with forward-only computations. In our framework this is in fact the case once we ignore decorations.

**Theorem 3 (Causal Consistency).** *If $\langle \sigma, \chi, \beta \rangle \in \mathsf{RC}_S$ and $\langle \sigma, \chi, \beta \rangle \rightsquigarrow \langle \sigma', \chi', \beta' \rangle$, then $\exists\, \langle \underline{\sigma}, \underline{\chi}, \underline{\beta} \rangle \in \mathsf{RC}_S$ such that* $\mathsf{forget}(\langle \sigma', \chi', \beta' \rangle) = \mathsf{forget}(\langle \underline{\sigma}, \underline{\chi}, \underline{\beta} \rangle)$.

## 7 Concluding Remarks and Related Work

We introduced a coordinated undo capability for CMFSs induced by a global graph. According to [8] our way of controlling reversibility is *semantic*, since the backward semantics of the machines, according to particular *guards* on the current configuration, decides when to revert the computation along a particular branch. Moreover, the semantics keeps track of the aborted/reversed branches in order to avoid executing again the path which led to the backward computation. Our framework offers some degree of freedom to easily tailor reversibility to different policies. Firstly, we deliberately left the logic to express reversion guards under-specified, opting for simplicity, for a propositional logic. Other options could be possible such as temporal logic. Another parameter of our framework is the reverse function $\rho$ that dictates how to revert a computation. Again, we here opted for simplicity and decided to undo the effects of one event only. Of course, other choices would also be reasonable; for instance, we contemplate to study variants where many (or all) reversible events are reversed at once.

Differently from fully reversible approaches [3, 13, 9] in our model input operations are potentially irreversible actions (unless they are into a loop) and backward transitions are confined only in branches. In [1] a contract model with rollbacks is presented. Like our approach, rollbacks reverts choices, and when a branch is reverted the model discards the failed branch. In our approach, the branch is not discarded, but disabled if the other branches can still be executed. If all the branches of a choice are disabled (i.e. all of them have been tried), then the choice behaves like a normal forward choice (e.g., the computations goes along one branch). Besides, in [1] reversibility is triggered when computation is stuck, and not upon the entailment of a condition and the model can express just interaction among two participants.

Global types [7] are another way to express a protocol of interactions from a global point of view. In [5] global types are extended with checkpoints: whenever a choice is traversed the system takes a *global* checkpoint and the system computation can be reverted to the last taken checkpoint. Our model is more expressive than [5], since we consider asynchrony and more importantly we do not have any restriction about the parallel operator. Indeed our running example cannot be expressed by [5]. Another difference with our work is that in [5] when a checkpoint is taken, also parties not related with the choice are forced to do a checkpoint. In other words, also parties not causally related with the choice are forced to participate to the checkpoint and eventually are forced to reverse their computation when the choice is reverted. In [12] global types are used to understand the potential recovery points of a protocol and to infer a partial causal order among messages. Our r-choreographies are more expressive than the global specification model in [12]; for instance, our running example cannot be represented with their global types (due to the parallel composition of graphs). Also, as discussed in [6], the causal conditions imposed on distributed choices by our semantics are less restrictive than the usual ones, of which the model of [12] is a representative. The greater expressiveness and the explicit handling of reversibility of loops require us to adopt a finer notion of run-time causality than the one adopted in [12]. As far as we understand, the only recovery that in [12] is done in presence of loops is possible only in the last iteration (past iterations do not seem to be reversible).

# A Running Example: the reversible CFSMs

The projections of the global graph of the running example into the participants T, B and D are respectively given in Fig. 4, Fig. 5 and Fig. 6.

For readability, we ignore control points in the transitions of the automata as messages in our running example unequivocally identify which interaction of the r-choreography in Example 2 they originate from. Also, in Fig. 4, the branching state $q_1$ is highlighted with a dashed line and we use

- dashed arrows for the transitions decorated with $q_1$, $\mathsf{T}{\cdot}\mathsf{B}!\mathsf{flight}$, $|\{\mathsf{upd} \in \mathsf{T}{\cdot}\mathsf{D} < 1\}|$,
- dotted arrows for the transitions decorated with $q_1$, $\mathsf{T}{\cdot}\mathsf{B}!\mathsf{car}$, $|\{\mathsf{upd} \in \mathsf{T}{\cdot}\mathsf{D} < 1\}|$, and
- dashed-dotted arrows for the transitions decorated with $q_1$, $\mathsf{T}{\cdot}\mathsf{B}!\mathsf{dest}$, $|\{\mathsf{upd} \in \mathsf{T}{\cdot}\mathsf{D} \geq 1\}|$.
- red dashed/dotted/dashed-dotted arrows which differs from their black version for having the decoration marked as brach exiting.

Finally, states represented with small circles are irrelevant in the following and therefore they are anonymous.

Consider the reachable configuration $\langle \sigma, \chi, \beta \rangle$ where

$$\chi(\mathsf{T}{\cdot}\mathsf{D}) = \varepsilon \ ; \ (\dagger, q_0\mathsf{T}, \mathsf{1}, 1), (\mathsf{upd}, q_{13}\mathsf{T}, \mathsf{1}, 4), (\dagger, q_{14}\mathsf{T}, \mathsf{9}, 5)$$
$$\chi(\mathsf{T}{\cdot}\mathsf{B}) = (\dagger, q_0\mathsf{T}, \mathsf{1}, 2), (\mathsf{dest}, q_1\mathsf{T}, \mathsf{8}, 3), (\dagger, _, 9, 6) \ ; \ \varepsilon$$
$$\chi(\mathsf{B}{\cdot}\mathsf{T}) = (\mathsf{fullP}, q_4\mathsf{B}, \mathsf{9}, 1) \ ; \ \varepsilon$$
$$\sigma(\mathsf{T}) = q_1 \quad \sigma(\mathsf{B}) = q_1 \quad \sigma(\mathsf{D}) = q_0 \qquad \text{and} \qquad \beta(q_1\mathsf{T}) = (_, \mathtt{ff})$$

Such configuration corresponds to an execution where T has executed the loop once asking B for a destination and has consumed the fullP message. Suppose now, that T decides to ask again for a destination dest. That is, the machine from state $q_1$ tries to reach state $q_4$. This transition is decorated with the guard $|\{\mathsf{upd} \in \mathsf{T}{\cdot}\mathsf{D} \geq 1\}|$ which is verified by the queue $\mathsf{T}{\cdot}\mathsf{D}$. This implies that we have to check whether log $\ell = (\mathsf{dest}, q_1\mathsf{T}, \mathsf{8}, 3) \in \mathtt{RBP}_{\mathsf{G},\chi}(\mathsf{T}{\cdot}\mathsf{B})$. According to Definition 7, since the message is on a loop (say $\mathsf{1}$) all its causes have to be on the same loop and the loop itself has to be ongoing. From the channels configuration we can see that loop $\mathsf{i}$ is still ongoing, and all its effects

$$(\mathsf{upd}, q_{13}\mathsf{T}, \mathsf{1}, 4), (\dagger, q_{14}\mathsf{T}, \mathsf{9}, 5)$$
$$(\dagger, _, _, 9)$$
$$(\mathsf{fullP}, q_4\mathsf{B}, \mathsf{9}, 1)$$

belong to the same loop. We then start by taking one of the logs caused by $\ell$ which has no causes, for example $(\dagger, q_{14}\mathsf{T}, \mathsf{9}, 5)$ and we set back the state of T machine to state $q_{14}$. The next minimal element is $(\dagger, _, _, 9)$ and so on. Eventually we reach a channels configuration of the form:

$$\begin{aligned}
&\chi'\big(\mathsf{T}\cdot\mathsf{D}\big) = \varepsilon\ ;\ \big(\dagger, q_0\mathsf{T}, \mathtt{1}, 1\big)\\
&\chi'\big(\mathsf{T}\cdot\mathsf{B}\big) = \big(\dagger, q_0\mathsf{T}, \mathtt{1}, 2\big)\ ;\ \varepsilon\\
&\chi'\big(\mathsf{B}\cdot\mathsf{T}\big) = \varepsilon\ ;\ \varepsilon\\
&\sigma(\mathsf{T}) = q_1 \quad \sigma(\mathsf{B}) = q_1 \quad \sigma(\mathsf{D}) = q_0
\end{aligned}$$

and branches $\beta\big(q_1\mathsf{T}\big) = (|\{\mathsf{upd} \in \mathsf{T}\cdot\mathsf{D} \geq 1\}|, \mathtt{tt})$. As desired.

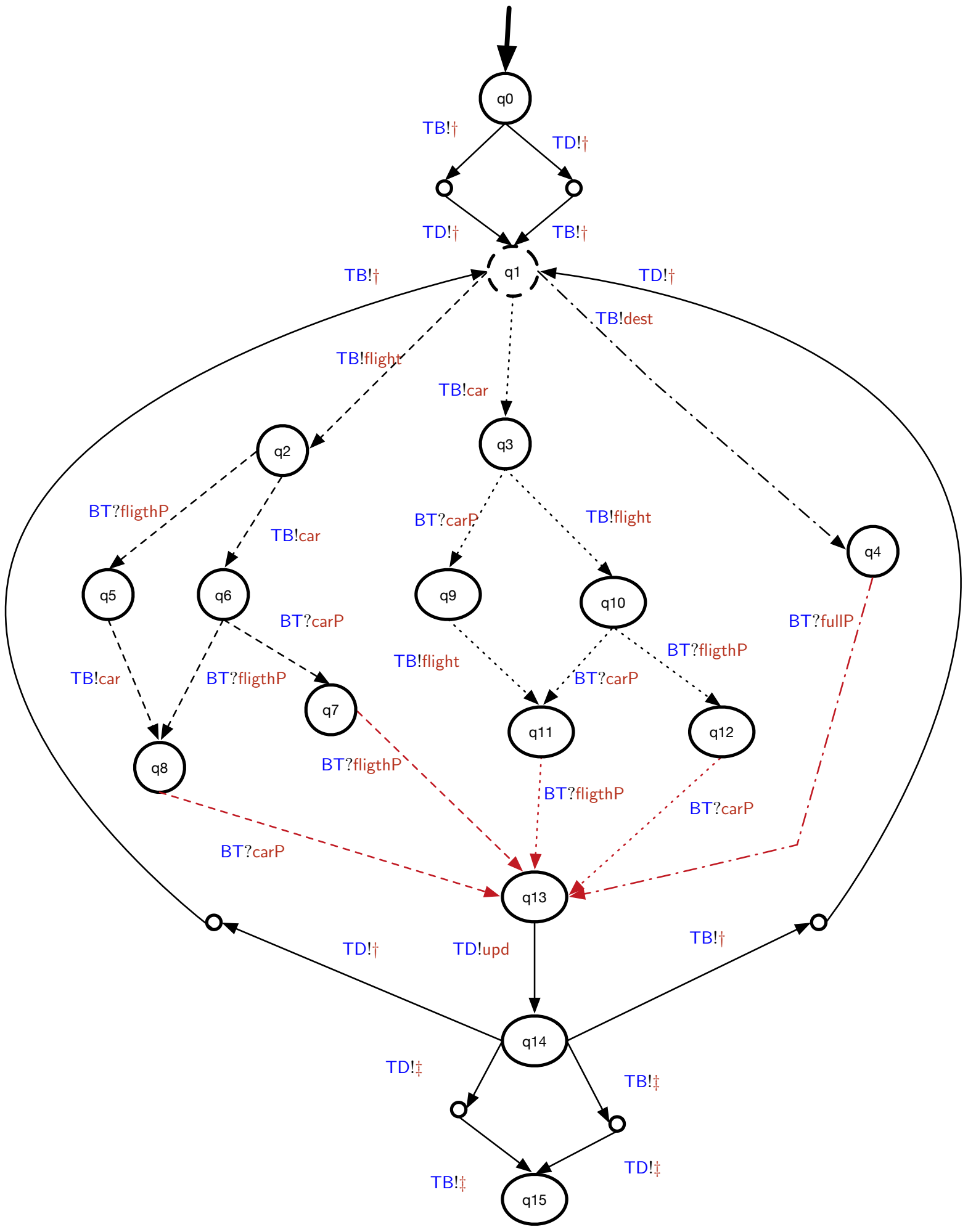

**Fig. 4.** Reversible CFSM for T

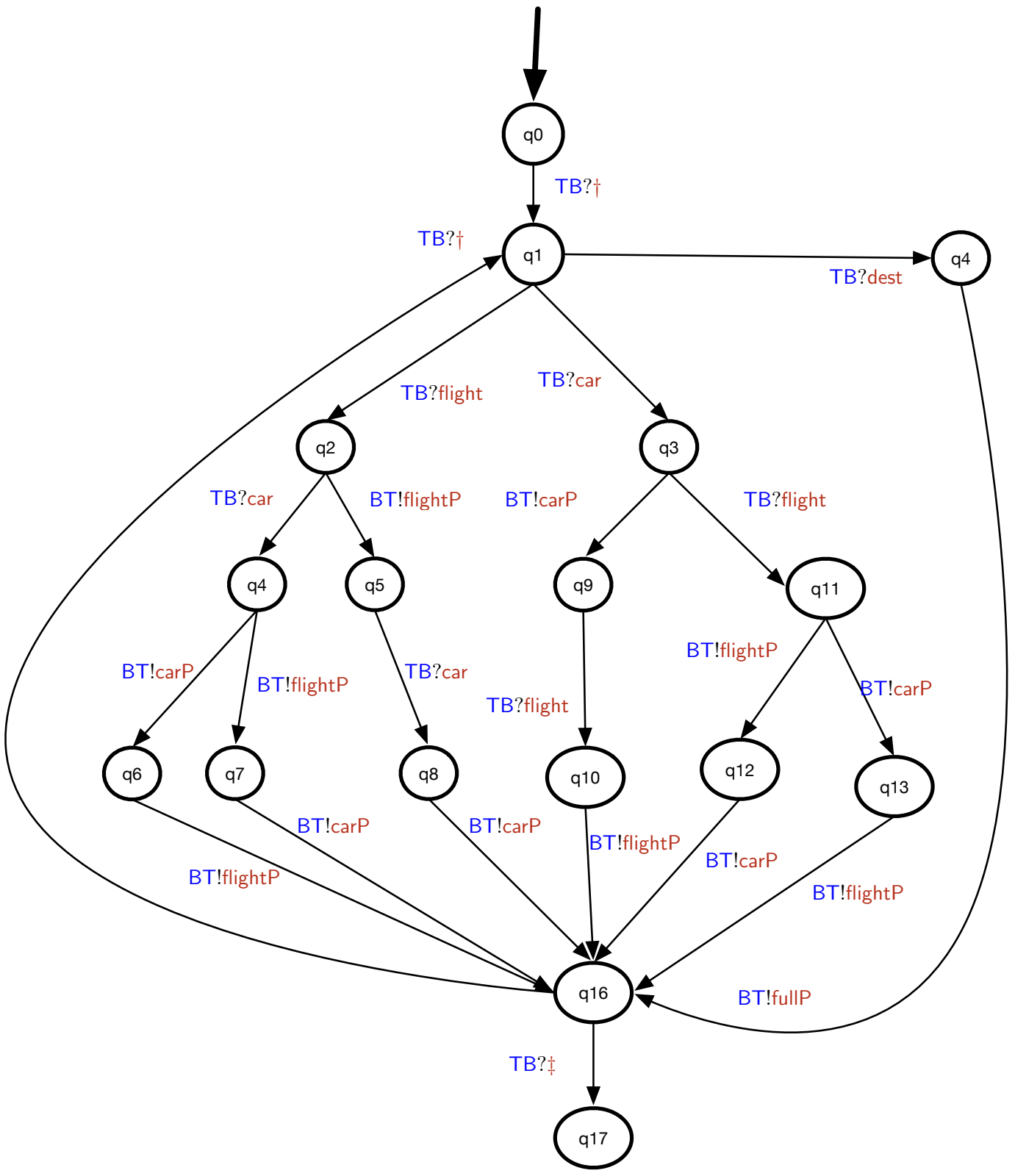

**Fig. 5.** Reversible CFSM for B

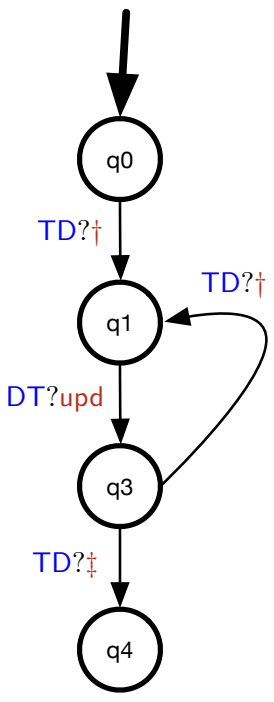

**Fig. 6.** Reversible CFSM for D